\documentclass[review]{tLCT2e}
\usepackage{graphicx}
\usepackage[FIGTOPCAP]{subfigure}

\newcommand{\PP}{\mathbf{P}}

\begin{document}


\title{Modulated nematic structures and chiral symmetry breaking in 2D}

\author{Karol Trojanowski, Micha\l{} Cie\'sla and Lech Longa\\\vspace{6pt} {\em{M. Smoluchowski Institute of Physics, Jagiellonian University, \L{}ojasiewicza 11, 30-248 Krak\'ow, Poland}};
}

\maketitle
\begin{abstract}
 We have studied the properties of biaxial particles  interacting via an anisotropic pair potential,  
 involving second rank
 quadrupolar and third rank octupolar coupling terms, using Monte Carlo simulation. The particles 
 occupy  the sites of a 
 2D square  lattice and the interactions are restricted to nearest neighbours. The system exhibits   
spontaneous chiral symmetry breaking  from an isotropic phase  to a chiral modulated nematic
phase,  composed  of  ambidextrous chiral domains.  When two-fold axes of  quadrupolar and 
octupolar tensors coincide  this modulated phase appears to be  the ambidextrous cholesteric 
phase of pitch  
comparable with a few lattice spacings, which can be regarded as a limiting case of the nematic twist bend 
phase.  The associated phase transition is first-order.
\end{abstract}
%
\section{Introduction}
Liquid crystal compounds, such as   chemically \emph{achiral}  flexible dimers 
\cite{PanovNTB,CestariNTB,BorshchNTB,ChenBananyPNAS},  trimers \cite{trimers} and   bent-core 
mesogens 
\cite{GleesonGoodby,ChenBananyPRE2014}  with their hybrids \cite{hybridbentcores},   can stabilize  
phases unlike anything recognized before. The most striking observation is one connected with
appearance  of  spontaneous chiral order, where domains of opposite optical activity are created in 
ordinary isotropic and nematic phases.  A new  macroscopically chiral
nematic  state that emerges,  known as the  twist-bend nematic  ($N_{TB}$), is  a
helicoidal structure with the director  tilted with respect to the helix
axis.  From the perspective of the purely nematic type of ordering  it can be considered as a 
generalization  of the 
cholesteric phase \cite{deGennes} where the  
director is orthogonal to  helix axis,  except for the observation of domains of opposite chirality and a 
short, nano-scale pitch, apparently spanning only
several  molecules.   This last property is particularly surprising  since in  the presently  
known cholesteric 
phase   the pitch is hundreds to thousands of molecules long.

In spite of its importance, the theory that correlates spontaneous  chiral symmetry breaking  (SCSB)
and modulated structures with relevant features of molecular interactions  is not yet fully 
developed and understood.  Taking into account  that the  pitch of $N_{TB}$  is so extremely short, it
seems likely that the driving force responsible for its formation  is the packing entropy connected with  
the bent shape (steric dipole) of the constituent molecules  \cite{2014Greco,2015prlGreco}.
Another possibility would be that chirality  could be  transmitted via
steric interactions emerging from the coupling to chiral conformations \cite{2016Ungar}.
It is this scenario that we would like to  discuss, at least partly, in the present paper.
\section{Chiral symmetry breaking in non-chiral materials}
At the phenomenological level  Dozov predicted the existence of  $N_{TB}$
as a result of the negative value of the bend elastic constant $K_3$  in the nematic 
phase \cite{DozovNTB}.  
 Since the resulting spontaneous bend cannot be extended globally
the uniform nematic phase  would become unstable
to the formation of a modulated phase, which could
be either chiral  $N_{TB}$, or nonchiral nematic splay-bend ($N_{SB}$).  
A possible  explanation of the sign change of the bend elastic constant is    
to assume that spontaneous local bend deformations of the
nematic director couple with emerging modulated polar order  via  the so called
flexoelectric effect  \cite{Meyer0,Meyer}. Indeed, as  follows from a direct calculation 
such polar order can effectively  reduce  $K_3$  and even make it negative 
\cite{SelingerNTB2013,lechNTB}.

But mesoscopic-level  consequences of the fact that the chiral symmetry breaking  
takes 
place in  the  nematic  phase are far more reaching.  To discuss some of  them let 
us assume that the \textit{primary order parameter} quantifying   
\textit{local} nematic order in $N_{TB}$   is a full  $3\times 3$ 
second-rank traceless and symmetric alignment tensor
field, $ {\mathbf{Q}}({\mathbf{r}})$  \cite{deGennes},  rather than its  director 
part only. In a standard parametrization $\mathbf{Q} $ can be written as
\begin{eqnarray}\label{eq:Q-diagonal}
\mathbf{Q} &=& {\frac{q_0}{\sqrt{6}}}\left(3\hat{\mathbf{n}}\otimes \hat{\mathbf{n}} -
{\boldsymbol{1}} \right) +
{\frac{q_2}{\sqrt{2}}}\left(\hat{\mathbf{l}}
\otimes \hat{\mathbf{l}} - \hat{\mathbf{m}}
\otimes \hat{\mathbf{m}}\right),
\end{eqnarray}
where
eigenvectors of $\mathbf{Q}$ are identified with the orthonormal, right-handed tripod  \{$\mathbf{\hat{l}},\mathbf{\hat{m}}, \mathbf{\hat{n}}$\}  of  
\emph{local}
directors,   corresponding to the \emph{local} eigenvalues 
$\lambda_1=-\frac{{q_0}}{\sqrt{6}}+\frac{{q_2}}{\sqrt{2}}$,
$ \lambda_2=-\frac{q_0}{\sqrt{6}}-\frac{q_2}{\sqrt{2}}
$,
$ \lambda_3=-\lambda_1-\lambda_2=\sqrt{\frac{2}{3}} q_0
 $,
respectively.  When three eigenvalues of $ {\mathbf{Q}}$
are equal, which gives $ {\mathbf{Q}}=0$,  the local structure is  an $SO(3)$-symmetric isotropic liquid. 
If $ {\mathbf{Q}}$ has two degenerate eigenvalues the local anisotropy is uniaxial, and biaxial if all three 
eigenvalues are distinct. These properties can be expressed using inequality between traces 
of $\mathbf{Q}^2$ and $\mathbf{Q}^3$ 
\begin{equation}\label{eq:traceRestriction}
    \mathrm{Tr}{(\mathbf{Q}}^{2}) ^3
     - 6 \mathrm{Tr}{(\mathbf{Q}}^{3}) ^2 \ge 0.
\end{equation}
The condition (\ref{eq:traceRestriction}) becomes
a strong inequality for locally biaxial (oblate or prolate)
configurations and is fulfilled as equality for the uniaxial ($\mathbf{Q} \ne \mathbf{0}$) and the isotropic (
$\mathbf{Q} = \mathbf{0}$) orientational ordering.  Excluding local isotropic configurations, the normalized 
parameter
\begin{equation} \label{eq:uniaxialityParameter}
-1 \le w = \frac{ \sqrt{6}\, \mathrm{Tr}\left(
                                   \mathbf{Q}^{3} \right)}{
   \left[\mathrm{Tr}\left( \mathbf{Q}^{2} \right) \right]^{\frac{3}{2}}
         } \le 1
\end{equation}
serves as  a scalar measure of how strongly uniaxial/biaxial is \textit{local} nematic order. 
For purely uniaxial phases $w^2$ is maximal and equals one while for phases of maximal 
biaxiality $w^2$ approaches its minimal value 0 \cite{longa-lt}.  

To link  SCSB with the local nematic order  a basic observation is that  a totally 
antisymmetric tensor $\varepsilon_{\alpha\beta\gamma}$, proportional to the Levi-
Civita tensor $\epsilon_{ijk}$, or equivalently, pseudotensor couplings between basic order parameters 
(responsible for SCSB) must spontaneously 
emerge at the transition to the chiral phase.  In the  lowest order scenarios, in addition to $\mathbf{Q}$, we need  at
least one more primary order parameter, which 
can be  either a first-rank vector field, say $\mathbf{P}(\mathbf{r})$, \cite{lechChiralFlexo} or a third-rank tensor 
field $\mathbf{T}(\mathbf{r})$, invariant with respect to 
tetrahedral  point group  symmetry \cite{FelTetrahedratic,bib:Lubensky2002}, or both 
\cite{bib:Lubensky2002}. The vector $\mathbf{P}$  could  
represent \textit{e.g.} mesoscopic 
polar order of steric and/or electric dipoles, or the wave vector  of the modulated structure. 
Non vanishing tensor  $\mathbf{T}$ would imply the 
presence of long-range octupolar part in third-rank nonlinear dielectric tensor. 
 In intrinsically chiral materials, where 
cholesteric and blue phases are stabilized, $\mathbf{T}$ and $\mathbf{P}$ fields can be correlated 
{\textit{e.g.}} with  $L=3$ and $L=1$ parts of $\partial _i Q_{j,k}$, respectively.
Lubensky 
and Radzihovsky  \cite{bib:Lubensky2002}  
have argued  that all three tensors $\mathbf{P}$,  $\mathbf{Q}$ and $\mathbf{T}$ are 
necessary to correctly account for  
symmetry breaking mechanisms observed in  bent-core  systems. 

As concerning SCSB, 
$\mathbf{P}$   and  $\mathbf{T}$  can both be used to construct the totally 
antisymmetric 
tensors $\varepsilon^{\mathbf{X}}_{\alpha\beta\gamma}$ (${\mathbf{X}}=\{\mathbf
{P},\mathbf{T}\}$), 
that provide a chirality measure for the emerging chiral structure. They 
are  given by
\begin{equation}\label{eq:pseudotensorP}
  \varepsilon^{\mathbf{P}}_{\alpha\beta\gamma}  =
  P_{[\alpha} (\mathbf{Q}\cdot\mathbf{P})_\beta (\mathbf{Q^2}\cdot\mathbf{P})_{\gamma ]}
  \propto |\mathbf{Q}|^3
  (\mathbf{\hat{l}\cdot P})(\mathbf{\hat{m}\cdot P})( \mathbf{\hat{n}\cdot P})  \\
   \times\sqrt{1-w^2}\,\, \epsilon_{\alpha\beta\gamma} = \varepsilon_{\PP} \,\epsilon_{\alpha\beta\gamma}
\end{equation}
\begin{equation} \label{eq:pseudotensorT}
   \begin{split}
  \varepsilon^{\mathbf{T}}_{\alpha\beta\gamma} = 
  Q_{ [ \alpha\mu} & (\mathbf{Q^2})_{\beta\nu} {T}_{\mu\nu\gamma ] }
  \propto |\mathbf{Q}|^3 |\mathbf{T}|
  \left[2(\mathbf{\hat{l}\cdot \hat{l}'})^2 + 2 (\mathbf{\hat{m}\cdot \hat{m}'})^2 \right. \\
    + 2( \mathbf{\hat{n}\cdot \hat{n}'})^2
   -6 &  \left. (\mathbf{\hat{l}\cdot \hat{l}'})(\mathbf{\hat{m}\cdot \hat{m}'})( \mathbf{\hat{n}\cdot \hat{n}'}) -1\right]
  \sqrt{1-w^2}\,\, \epsilon_{\alpha\beta\gamma}=  \varepsilon_{\mathbf{T}} \,\epsilon_{\alpha\beta\gamma}.
  \end{split}
\end{equation}
Here $[ ...]$  denotes antisymmetrization over indices $\alpha, \beta, \gamma$; $\{\mathbf{\hat
{l}', \hat{m}', \hat{n}'}\}$ is the 
orthonormal tripod of vectors, 
parallel to 2-fold rotation 
axes of the octupolar tensor $\mathbf{T}$; $|\mathbf{Q}|=\sqrt{Q_{\alpha\beta} Q_{\alpha\beta}}$ 
and  $|\mathbf{T}|= \sqrt{T_{\alpha\beta\gamma} T_{\alpha\beta\gamma}} $. In addition we have
used 
an Einstein summation convention for the repeated indices. The primary conclusion from the Eqs.
(\ref{eq:pseudotensorP},\ref{eq:pseudotensorT}) is that SCSB  in nematics, irrespective of 
the way
it is realized in practise,  should stabilize a structure which is described  \textit{locally} by the   
\textit{biaxial} tensor field $\mathbf{Q}$. We should mention that the necessity of considering  the full biaxial field 
$\mathbf{Q}$ for a proper understanding of  SCSB, rather than its uniaxial part only,
is in line with the observation that all chiral phases of at least
intrinsically chiral mesogens are  biaxial \cite{lechChiralIcosa}. For cholesterics of 
periodicity being 
of  the order of 500 nm  this  biaxiality can be  weak and homogeneous  ($w\approx 1$),  but it 
becomes relevant  for blue phases, where $w$  being space-dependent, varies between -1 and 1. 

The condition that biaxiality of $\mathbf{Q}$ is necessary to induce chiral order is, however,
not sufficient. For example, for a vector fields $\PP$ to contribute to chirality measure   $\varepsilon_{\PP}$  we need, 
in addition, that $\PP$ does not belong to a plane spanned by any two vectors of  
the tripod \{$\mathbf{\hat{l}},\mathbf{\hat{m}}, \mathbf{\hat{n}}$\}.
From the formulas (\ref{eq:pseudotensorP},\ref{eq:pseudotensorT}) one can further conclude that the coefficients $\varepsilon_{\PP}$ and $\varepsilon_\mathbf{T}$ measuring sign and 'degree of chirality' for given $\mathbf{Q}$, $\mathbf{P}$ and $\mathbf{T}$ are restricted by inequalities
\begin{eqnarray} \label{ep}
-1 \le &\frac{3 \sqrt{3} \,\varepsilon_{\PP}}{|\mathbf{Q}|^3  |\mathbf{P}|^3 \sqrt{1-w^2}}&\le 1  \\ \label{et}
-1\le &\frac{\varepsilon_\mathbf{T}}{|\mathbf{Q}|^3  |\mathbf{T}| \sqrt{1-w^2}}&\le 1.
\end{eqnarray}
Thus, the extremal  value of  chirality can be  achieved  for  
\begin{equation}
\mathbf{P}=\frac{|\mathbf{P}|}{\sqrt{3}} (\pm \mathbf{\hat{l}\pm\hat{m}  \pm\hat{n}}),
\end{equation}
where odd number of  $`-`$ signs   corresponds 
to a state from  the  upper  limit in (\ref{ep}) while the remaining  combinations of  $`+`$ and $`-`$ signs are states from
the  lower  bound  in (\ref{ep}). Likewise, taking 2-fold axes   of  $\mathbf{Q}$  parallel to 2-fold axes   of  $\mathbf{T}$  
\begin{equation}  \label{minmax}
 \{\mathbf{\hat{l}},\mathbf{\hat{m}}, \mathbf{\hat{n}}\} \,\, ||\,\,
 \{\pm \mathbf{\hat{l}'},\pm\mathbf{\hat{m}'}, \pm\mathbf{\hat{n}'}\}
\end{equation}
 gives states satisfying lower bound in (\ref{et}), while permutation of any two  vectors on the right-hand side 
 of  (\ref{minmax}) corresponds to upper bound states. The allowed choice between  $`+`$ and $`-`$ 
 signs in   
 (\ref{minmax})  is such that  the handedness of  both bases  should be the same.
 
It is likely that  for  molecular systems  exhibiting SCSB  both ways of 
acquiring structural chirality can be important. But except for symmetry classification of  
spontaneous order \cite{bib:Lubensky2002}  only special, separate 
cases of  \{$\mathbf{Q}$,$\mathbf{P}$\} and  \{$\mathbf{Q}$ ,$\mathbf{T}$\} couplings have been
discussed in the literature. Starting from a formal theory of flexopolarization  
for systems described in terms of  \{$\mathbf{Q}$,$\mathbf{P}$\}  \cite{lechChiralFlexo} 
possible equilibrium one-dimensional modulated structures were identified, both for nonchiral and intrinsically chiral materials  \cite{lechNTB}.
The theory permits stabilization of $N_{SB}$, a few variants of $N_{TB}$ -from weakly to strongly biaxial-   
and a new class of one-dimensional achiral modulated nematic structures. 
More complex structures, like polar 2D hexagonal  and 3D bcc analogues of blue phases, can  
also form \cite{SelingerPolarBluePhases2014}. 

Couplings between  the pair of  \{$\mathbf{Q}$,$\mathbf{T}$\} fields have
been shown to generate even  larger  class of new phases, ranging
from  the tetrahedratic liquid to novel
biaxial, polar, and chiral phases \cite{bib:Lubensky2002,Brand2005189,Brand2010}. These 
phenomenological predictions are consistent  with a few molecular level
studies.  In particular, Bisi et al. \cite{doi:10.1080/15421401003795670} showed that 
rigid $C_2$-symmetric molecules generate 
quadrupolar and octupolar terms to the
Onsager`s excluded volume, which is prerequisite for having
biaxial and tetrahedratic ordering in the mean-field theory.
Evaluation of point dispersion interactions \cite{DF9654000232} between two
bent-core molecules gives mathematically similar terms \cite{bib:Longa2009}.
Translating this to microscopic interactions we have studied a class of 3D,  generalized Lebwohl-Lasher  lattice 
dispersion models \cite{bib:Longa2009,bib:Trojanowski2012,bib:Longa2013} with nearest-neighbour 
interactions involving quadrupolar and octupolar couplings.  We considered only the maximal chirality 
model  (MCM), where two-fold axes of   $\mathbf{Q}$ and $\mathbf{T}$  coincide (\textit{see} discussion after formula (\ref{et})).
Both, molecular-field calculations and Monte Carlo computer simulations  
proved   the formation
of absolutely stable tetrahedratic, tetrahedratic nematic, and
chiral tetrahedratic nematic liquids of global $T_d$ , $D_{2d}$ , and $D_2$
symmetry, respectively, in addition to the standard uniaxial and
biaxial nematic phases.
 Here we carry out Monte-Carlo simulations for the model  \cite{bib:Trojanowski2012} in 2D.   
 We show that the model exhibits 
 spontaneous chiral symmetry breaking from an isotropic phase ($I$)  to a 
 nematic 
 twist-bend-like structure, which appears to have a nanoscale cholesteric 
 arrangement.  The corresponding phase transition appears to be first-order.
\section{Model}
We consider the dispersion interaction potential  which  accounts,  in an averaged way,  for 
chirality that can be
induced \textit{e.g.} by conformational degrees of freedom. It is defined by a  coupling between 
molecular
multipole 
moments that have  quadrupolar and octupolar parts. More specifically, we assume each molecule to 
consist of the point  quadrupolar moment $\mathbf{Q}( \mathbf{\hat{\Omega}} )$  and 
the molecular octupolar moment 
$\mathbf{T^{(3)}_2}( \mathbf{\hat{\Omega}} )$, which is a spherical, third-rank tensor 
of $T_d$ point group 
symmetry. The  quadrupolar tensor is given by
\begin{equation}
\mathbf{Q}( \mathbf{\hat{\Omega}}) = \mathbf{T^{(2)}_0}( \mathbf{\hat{\Omega}} ) + \lambda \sqrt{2} \mathbf{T^{(2)}_2}( \mathbf{\hat{\Omega}} )
\end{equation}
with the  $D_{\infty h}$-symmetric (uniaxial) part 
\begin{equation}
\mathbf{T^{(2)}_0}( \mathbf{\hat{\Omega}}) = \sqrt{\frac{3}{2}}\left( \mathbf{\hat{c}} \otimes \mathbf{\hat{c}} - \frac{1}{3}{\mathbb I} \right)
\end{equation}
and the $D_{2h}$-symmetric biaxial part 
\begin{equation}
\mathbf{T^{(2)}_2}( \mathbf{\hat{\Omega}} ) = \frac{1}{\sqrt{2}} \left( \mathbf{\hat{a}} \otimes \mathbf{\hat{a}} - \mathbf{\hat{b}} \otimes \mathbf{\hat{b}} \right).
\end{equation}
The right-handed  molecular basis $ \mathbf{\hat{\Omega}}=(\mathbf{\hat{a}},\mathbf{\hat{b}},\mathbf{\hat{c}})$ 
is specified by three orthonormal vectors firmly attached to the molecule and  
fixed  parallel to two-fold axes of $\mathbf{Q}$
and $\mathbf{T^{(3)}_2}$. That is, the configuration of multipoles corresponds to MCM introduced in the previous section (\ref{et}).  
Parameter $\lambda$ controls molecular biaxiality. In particular,  Eq.~
(\ref{eq:uniaxialityParameter}) now reads
\begin{equation}\label{wmol}
w=\frac{1-6 \lambda ^2}{\left(2 \lambda ^2+1\right)^{3/2}}.
\end{equation}
The eigenvalues of $\mathbf{Q}$ corresponding to 
   eigenvectors $(\mathbf{\hat{a}},\mathbf{\hat{b}}, \mathbf{\hat{c}})$ are  $\left(\lambda -\frac{1}{\sqrt
{6}},-\lambda
   -\frac{1}{\sqrt{6}},\sqrt{\frac{2}{3}}\right)$, 
   respectively. 

The octupolar ($T_d$-symmetric) tensor reads
\begin{equation}\label{t32}
\mathbf{T^{(3)}_2}( \mathbf{\hat{\Omega},p} ) =\frac{p}{\sqrt{6}} \sum_{\mathbf{(\hat{x}},\mathbf{\hat{y}}, \mathbf{\hat{z})} \in \pi(\mathbf{\hat{a}}, \mathbf{\hat{b}}, \mathbf{\hat{c})}} \mathbf{\hat{x}} \otimes \mathbf{\hat{y}} \otimes \mathbf{\hat{z}} ,
\end{equation}
where summation runs over all permutations of the molecular basis and where  $p=\pm 1$   is the parity degree of freedom \cite{bib:Trojanowski2012}. We can correlate  $p$  with \textit{e.g.} conformational chirality by  noting that  all chemically \emph{achiral}  molecules  that form stable $N_{TB}$ can be found in  chiral configurations which  fluctuate equally between positive and negative
chiralities.  
To account for  such  chirality fluctuations   we need at least one  
additional degree of freedom  per  molecule, here denoted   $p$,  which allows to distinguish between
the two classes of molecular conformations of opposite chirality. To both of these classes there
may correspond many conformational states of a molecule, but we assume that they do not modify the norm and relative orientation of quadrupolar and octupolar moments  of each class 
in an essential way.

For the model considered the two  opposite chiralities  are realized when reflections of the molecular basis
that do not preserve handedness  are included. A parity flip amounts to inverting one or more of the molecular axes or,
equivalently, changing the sign of  $\mathbf{T^{(3)}_2}$.   It modifies  $\mathbf{T^{(3)}_2}$
in accordance with  Eq.~(\ref{t32}), but leaves  $\mathbf{Q}$   unaffected.

\begin{figure}[htb]
\centerline{%
\includegraphics[width=0.7\columnwidth]{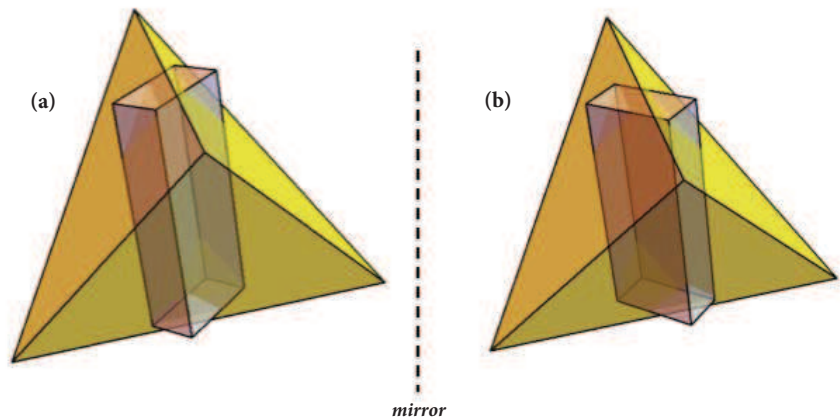}
}
\caption{
Maximal chirality model represented by a combination of tetrahedron and cuboid  (a), where  
their two-fold symmetry axes  are  kept  parallel to each other. Note that  mirror image (b) of (a)   cannot be brought into coincidence with (a)  using  translations and proper rotations. When  tetrahedrons are made to coincide, cuboids are misaligned and \textit{vice versa}. }
\label{fig:symmetries}
\end{figure}

The role  of  $\mathbf{Q}$  and $\mathbf{T^{(3)}_2}$ multipoles in chiral symmetry breaking   is  further illustrated in   Fig.\ref{fig:symmetries} where symmetry of   $\mathbf{Q}$   is represented by a cuboid   and   that of $\mathbf{T^{(3)}_2}$   by  a tetrahedron.    
Separately, 
cuboid and tetrahedron are  achiral, because  cuboid (tetrahedron) and its mirror image  can be superimposed by applying a  translation and  a proper rotation. However, when they are coupled by fixing mutual orientation of their two-fold axes this is  not true, in general. Clearly, for the case of  MCM,  where two-fold axes are kept parallel as shown in  Fig.\ref{fig:symmetries},   the configuration is chiral.  But, if we replace the cuboid with a cylinder, representing uniaxial symmetry,  the chirality is lost, in agreement with  Eq.~(\ref{eq:pseudotensorT}).
Taking all the above considerations into account we can now construct  a lattice dispersion model that allows to study  a possibility of  chirality fluctuations due to { \textit{e.g.}} conformational degrees of freedom.  It is  defined by the Hamiltonian \cite{bib:Longa2009,bib:Trojanowski2012,bib:Longa2013}
\begin{equation} \label{hamiltonian}
H=\frac{1}{2} \sum_{<i,j>}^N \left[ 
V( p_i , \mathbf{\hat{\Omega}_i}, p_j , \mathbf{\hat{\Omega}_j} ) + V_c(p_i, \mathbf{\hat{\Omega}
_i}, p_j, \mathbf{\hat{\Omega}_j} ) 
\right].
\end{equation}
The sum in (\ref{hamiltonian}) runs over nearest neighbours, i.e. for each particle i the sum runs over z = 6
neighbouring particles j, while the N particles occupy sites of a simple cubic lattice with
periodic boundary conditions. The arguments  $\mathbf{\hat{\Omega}_i}$
and  $\mathbf{\hat{\Omega}_j}$  denote  the right-handed molecular
frames of reference for molecules $i$ and $j$, respectively, while $p_i=\pm 1$ and $p_j=\pm 1$  take into account  parity of the molecular conformations.

The pair dispersion potential between neighbouring molecules  involves two terms. The first one  is
given by 
\begin{equation}\label{Vqqtt}
V(p_i, \mathbf{\hat{\Omega}_i}, p_j, \mathbf{\hat{\Omega}_j} ) = -\epsilon \left[ \mathbf{Q}( \mathbf{\hat{\Omega}_i} ) \cdot \mathbf{Q}( \mathbf{\hat{\Omega}_j} ) + \tau  \mathbf{T^{(3)}_2}(p_i,  \mathbf{\hat{\Omega}_i} ) \cdot \mathbf{T^{(3)}_2}(p_j,  \mathbf{\hat{\Omega}_j} ) \right].
\end{equation}
It represents the interaction between quadrupolar  \cite{doi:10.1080/00268977500102881}  and octupolar 
\cite{FelTetrahedratic} moments at sites $i$ and $j$, where
the scalar product $`\cdot`$
is understood as a full contraction over Cartesian indices.

The second term, denoted  $V_c$ , represents the lowest order coupling involving  quadrupolar and  octupolar  moments, 
and the intermolecular unit vector  $\mathbf{\hat{r}}_{ij}$. It  reads \cite{bib:Trojanowski2012}
\begin{equation}
V_c(p_i, \mathbf{\hat{\Omega}_i}, p_j, \mathbf{\hat{\Omega}_j} ) = \frac{\kappa}{\epsilon}\left[ {\Theta}_{\alpha \beta \gamma}( \mathbf{\hat{\Omega}_i} ) {Q}_{\alpha \nu}( \mathbf{\hat{\Omega}_i}) {Q}_{\beta \nu}( \mathbf{\hat{\Omega}_j} ) -
{\Theta}_{\alpha \beta \gamma}( {\hat{\Omega}_j}) {Q}_{\alpha \nu}( \mathbf{\hat{\Omega}_j}) 
{Q}_{\beta \nu}( \mathbf{\hat{\Omega}_i} )
 \right] \left( \mathbf{\hat{r}_{ij}} \right)_{\gamma},
\end{equation}
 where   $\Theta_{\alpha \beta \gamma}$
is given by
\begin{equation}
\label{Vc}
{\Theta}_{\alpha \beta \gamma}(p_i, \mathbf{\hat{\Omega}_i}) = 2\sqrt{2} 
\sum_{({{x}},{{y}}, {{z}}) \in c({\alpha}, {{\beta}}, {{\gamma}})} T^{(2)}_{0,x\mu}( \mathbf{\hat{\Omega}_i})
T^{(2)}_{2,y\nu} ( \mathbf{\hat{\Omega}_i}) T^{(3)}_{2,\mu\nu z} (p_i,  \mathbf{\hat{\Omega}_i}).
\end{equation}
Here summation runs over cyclic permutations $c(\alpha,\beta,\gamma)$ of $\{ \alpha,\beta,\gamma\}$. 
 It is perhaps  worthwhile to add that   
terms similar to (\ref{Vqqtt}-\ref{Vc}) can be generated by considering
multipole expansion of  the 
Onsager's excluded volume between two 
bent-core molecules in chiral conformations. The only difference would be 
additional  polar couplings, which we disregarded here. 

Special cases of the model  (\ref{hamiltonian})  have already been studied. For $\tau=\lambda=\kappa=0$  the
model  reduces to  the  Lebwohl-Lasher \cite{PhysRevA.6.426} potential,
which accounts for isotropic and uniaxial nematic phases
connected by a first-order phase transition.  Luckhurst \textit{et al.} 
considered  $\tau=\kappa=0$ case \cite
{doi:10.1080/00268977500102881,doi:10.1080/00268978000101341}
with nonzero $\lambda$ parameter, which controls
the biaxiality of a molecule. 
For $0\le\lambda \le \sqrt{\frac{3}{2}}$   the  parameter $w$, Eq.~(\ref
{wmol}),   covers the whole interval 
of allowed values approaching maximal biaxiality case ($w=0$)  at the so called self-dual point  ( $\lambda=1/\sqrt[]{6}$), 
where molecules are neither prolate nor oblate. The self-dual point separates phases in which  
the biaxial molecules are  prolate-like
 ($\lambda <1/\sqrt[]{6}$)  from  phases in which the molecules are
oblate-like  ($\lambda >1/\sqrt[]{6}$),  while
the boundary cases correspond to Maier-Saupe-like uniaxial models for long rods $(\lambda= 0)$
and platelets $(\lambda=\sqrt{\frac{3}{2}}) $.  The
phase diagram for varying  $\lambda$ has been obtained using mean-field theory and confirmed
by Monte Carlo simulations \cite{PhysRevLett.75.1803}.
The model predicts  a prolate uniaxial
nematic  phase,  an oblate uniaxial  nematic  phase, a biaxial nematic 
phase, and an isotropic  phase, where  a
sequence of a first-order transition to the uniaxial nematic and second-order transition
to the biaxial nematic  occurs with lowering temperature. At  $\lambda=1/\sqrt[]{6}$ only a direct
transition from the isotropic  to the biaxial nematic phase takes place.

When only the octupolar coupling term, proportional to $\epsilon \tau$
is retained in Eq. (\ref{Vqqtt}), the model predicts first-order transition from
the isotropic  phase to the tetrahedratic  phase of $T_d$ symmetry and 
 was studied by  Romano \cite{PhysRevE.77.021704} and
by one of us \cite{bib:Longa2009,bib:Trojanowski2012}.
A  full  analysis of the model when $\kappa=0 $   is given in  \cite{bib:Longa2009,bib:Trojanowski2012}. We 
identified, in addition to  the uniaxial,  biaxial and tetrahedratic nematic phases, 
two further spatially homogeneous nematic like phases. They involved 
nematic tetrahedratic   and  chiral nematic tetrahedratic  phases of global  $D_{2d}$, and $D_2$ symmetry, 
respectively.

For  the most complex case, where additionally  $\kappa\ne 0$,  only  preliminary  results  are available  
\cite{bib:Trojanowski2012,bib:Longa2013}.  Such coupling, as
we showed, can superimpose spatially inhomogeneous (short or
long-range) orientational order on the structures already
identified. In particular, the chiral nematic tetrahedratic  phase becomes unstable against
spontaneous twist formation. Indeed, by considering the minimum of the interaction potential for two isolated neighbouring molecules  we find that it favours a locally twisted configuration 
with  pitch being of the order of  $\pi (\Lambda+4\tau)/(\kappa \Lambda\overline{p}) $, where $\Lambda=3 + 2\lambda (\sqrt[]{6}+\lambda)$  and where  $\overline{p}$  is 
the average 
parity  ($\overline{p}=1$ or $\overline{p}=-1$  for biaxial molecules in the ground state)\cite{bib:Trojanowski2012}. In other words, both for   phases with $\overline{p}\ne 0$  and with  $\overline{p}=0$,    twisted
domains of opposite handedness will form in equal abundance, leading to
ambidextrous chirality.  

\begin{figure}[htb]
\centerline{%
\includegraphics[width=0.7\columnwidth]{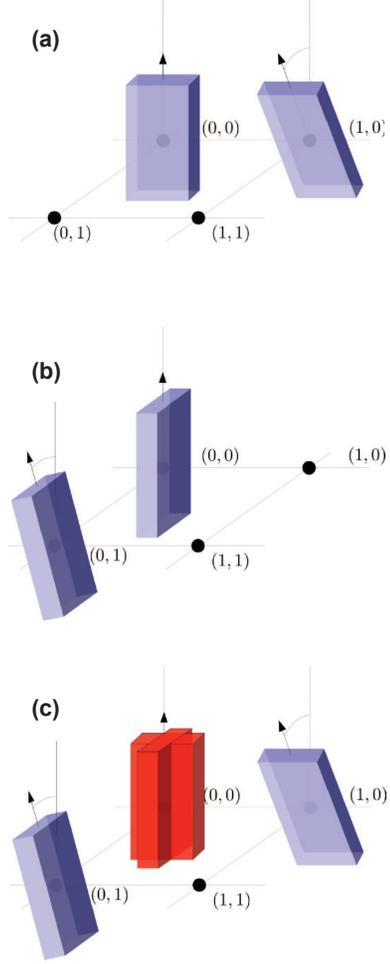}
}
\caption{Illustration of how frustration emerges for $\kappa\ne 0$ (\ref{Vc}) when three instead
of two biaxial  molecules with identical parity $(p = +1)$ are considered. Molecules are represented
by  cuboids, which roughly correspond to the molecular quadrupolar tensors  $\mathbf{Q}(\mathbf{\hat{\Omega}})$:
a) Ground state of two isolated molecules placed on neighbouring lattice sites along  $\mathbf{\hat{x}}$.
b) Ground state of two isolated molecules placed on neighbouring lattice sites along  $\mathbf{\hat{y}}$.
c) The two pairwise ground states (a) and (b) conflict each other.  
 }\label{fig:frustration}
\end{figure}

An important phenomenon associated with a formation of spontaneous twist in the model
(\ref{hamiltonian})
is \textit{frustration} of orientational 
order. This is illustrated in  Fig.~(\ref{fig:frustration}), where three biaxial molecules are placed in the
($\mathbf{\hat{x}},\mathbf{\hat{y}}$)  plane  with the intermolecular vectors constrained to four directions 
$\mathbf{\hat{r}}_{ij} =
\{1, 0\}, \{0,1\}, \{-1,0\}, \{0,-1\}$.
Consider now that two molecules, 1 and 2, of identical parity $p_1 = p_2 
= +1$, occupy
neighbouring sites at positions $(x_1, y_1) = (0, 0)$ and $(x_2, y_2) = 
(1, 0)$ (Fig. \ref{fig:frustration}a).
As
discussed above, the ground state of such configuration is achieved
when both molecules 1 and 2 align with $\mathbf{\hat{a}}_1 \parallel 
\mathbf{\hat{a}}_2\parallel
\mathbf{\hat{r}}_{12} = \{1, 0\}$  and molecule 2 is tilted
clockwise with respect to 1 around $ \mathbf{\hat{a}}_2$ (or vice-versa). 
Now consider a third molecule, 3,
$p_3 = p_1 = p_2 = +1$, located at $(x_3, y_3) = (0, 1)$. If 1 and 3 were 
treated in isolation  (Fig. \ref{fig:frustration}b), their ground state configuration
would be achieved analogously, by aligning $\mathbf{\hat{a}}_3 \parallel 
\mathbf{\hat{a}}_1\parallel
\mathbf{\hat{r}}_{13} $ and
tilting 3 clockwise with respect to 1 along $\mathbf{\hat{a}}_3$ (or vice-
versa). Now notice that the two
pairwise ground states for 1 and 2 and for 1 and 3 cannot be achieved 
simultaneously
(Fig. \ref{fig:frustration}c). Adding more molecules multiplies the 
number 
of conflicting conditions.
Thus, the system is frustrated. The question of how the system can relax  
frustrated configurations
is difficult to answer on analytical grounds and in the present we resort 
to Monte Carlo simulation in
search for an answer in 2D. The 3D case is postponed to our forthcoming 
studies.  
We demonstrate that the 2D MCM model releases frustration by a first-
order, 
isotropic- ambidextrous cholesteric phase transition.  The obtained ambidextrous 
cholesteric phase ($N^*_A$)
can be regarded as the limiting case of $N_{TB}$, where the director tilts at right angle with 
respect to the helix axis.
\section{Results and conclusions}
We perform  Monte-Carlo simulations of a system of molecules
interacting 
with 
the hamiltonian (\ref{hamiltonian}) on a two dimensional, square 
grid with periodic boundary conditions in the lattice plane and free
boundary 
conditions in the 
perpendicular direction. Two different sample sizes are used: $32 
\times 32$ and $64 \times 64$ to ensure 
that finite size effects are kept under the control. We take  $\kappa 
= \tau = 1.0$, $0.1 \lesssim \lambda \lesssim 0.9$ and use the 
dimensionless, 
reduced temperature $t=k_BT/
\epsilon$ for temperature scans. If not stated othervise simulations for different temperatures are
initialized from a random, disordered state.
In each Metropolis step a random site in the grid is selected.  Each attempted
MC move involves proper
random rotation (generated using quaternions)  of the molecular frame and 
parity inversion. Whenever possible, the size of the MC rotational step 
is 
adjusted to give
acceptance ratio at the level of  $0.3-0.5$.
  
The thermalization process is system's size dependent and lasts, on the 
average, for about $10^5$ cycles. After 
thermalization, the production run involves $10^5$ cycles with every 
tenth cycle configuration taken to calculate thermodynamic averages.
\begin{figure}[htb]
\centerline{%
\subfigure[$t=0.9$]{\includegraphics[width=0.4\columnwidth]{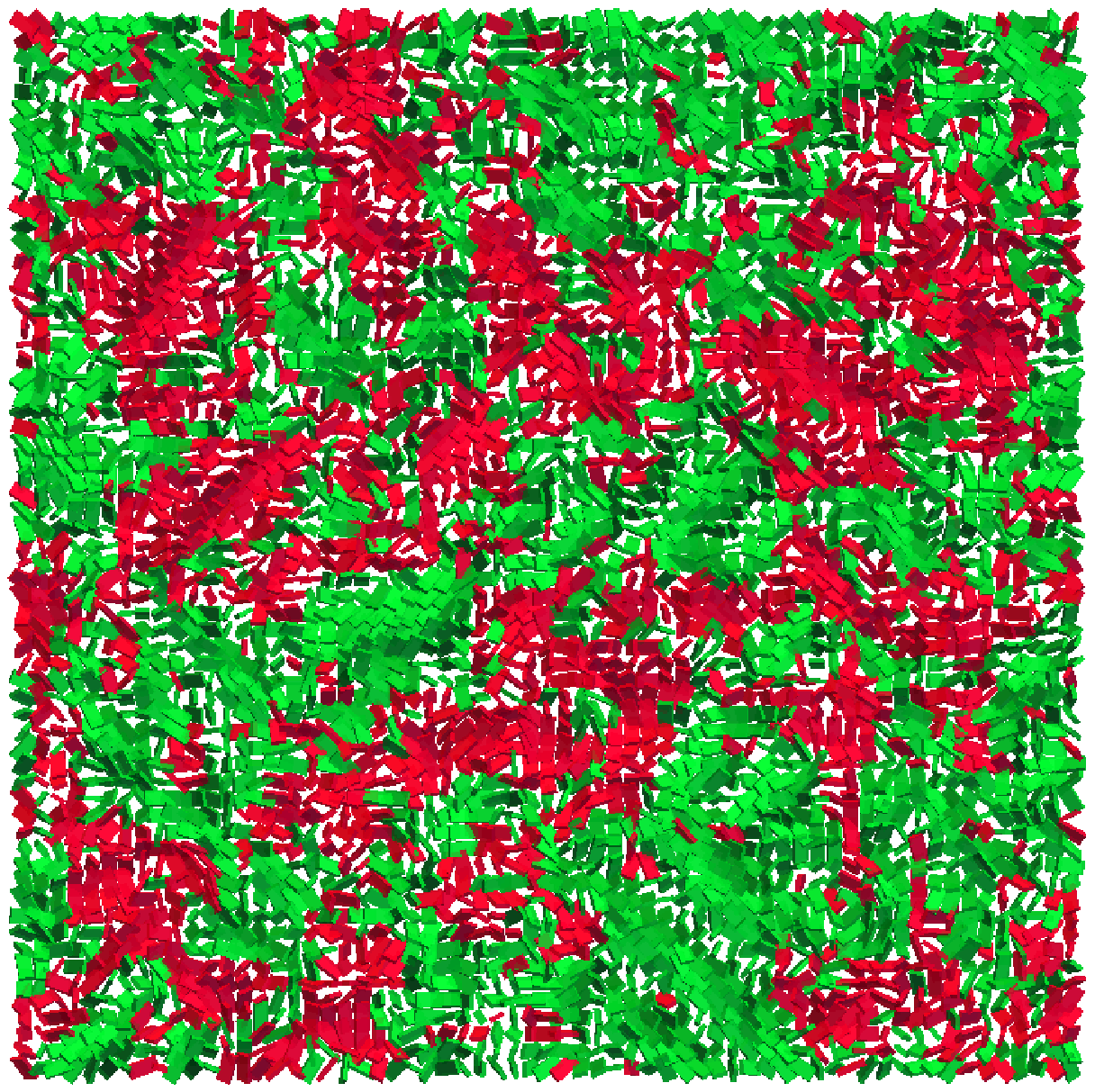}}
\hspace{0.01\columnwidth}
\subfigure[$t=0.896$]{\includegraphics[width=0.4\columnwidth]{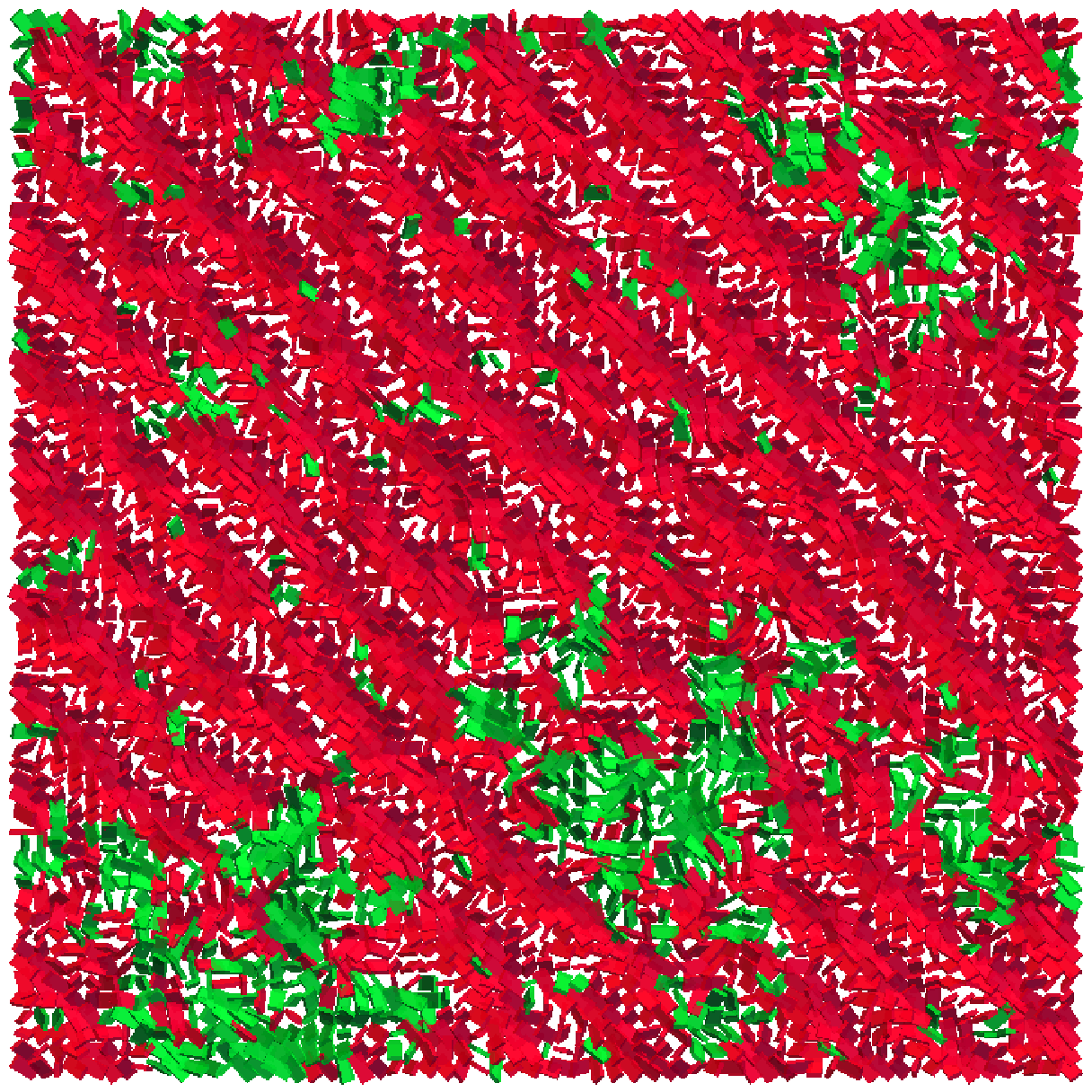}}
}
\vspace{0.01\columnwidth}
\centerline{%
\subfigure[$t=0.850$]{\includegraphics[width=0.4\columnwidth]{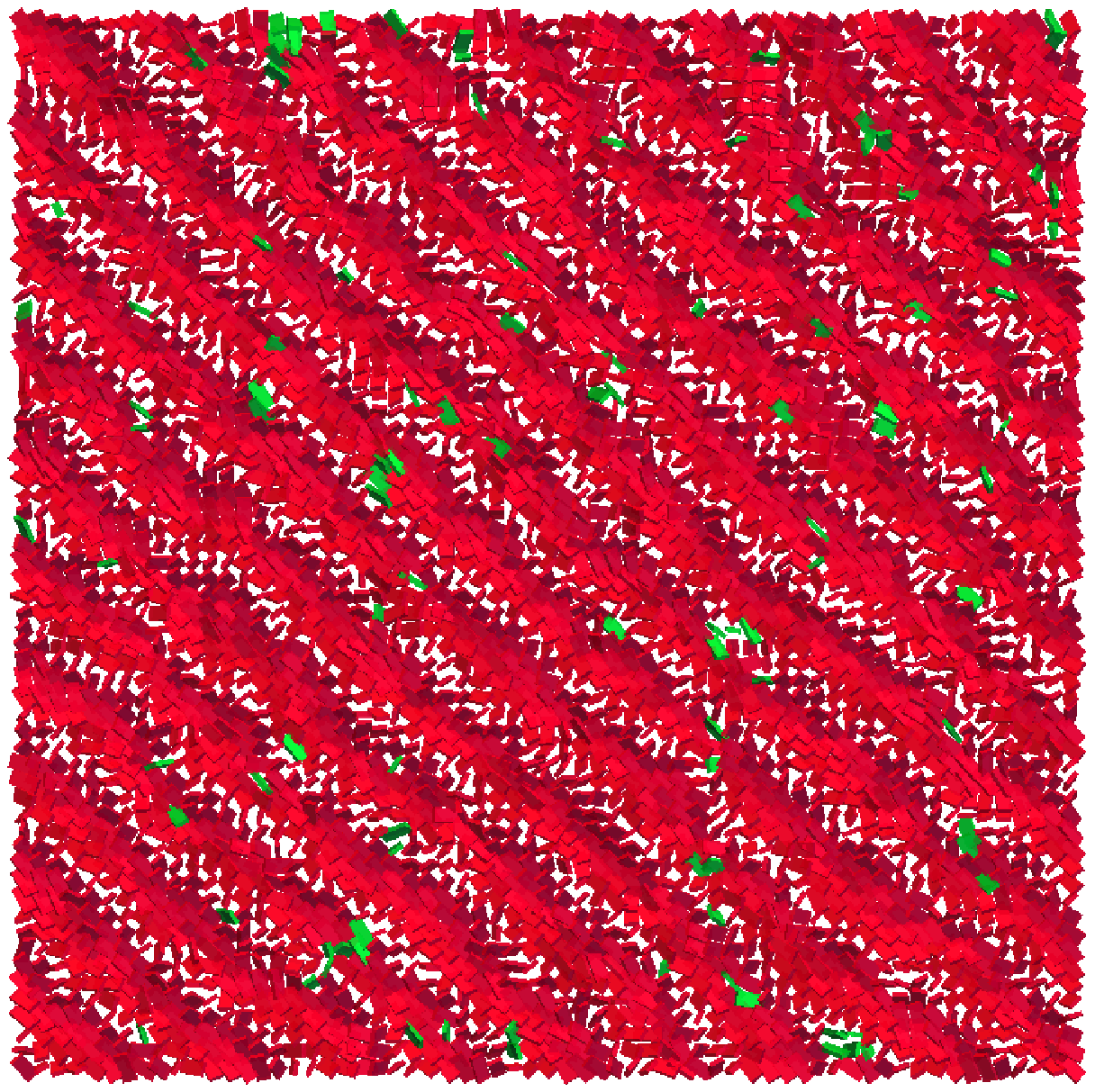}}
\hspace{0.01\columnwidth}
\subfigure[$t=0.4$]{\includegraphics[width=0.4\columnwidth]{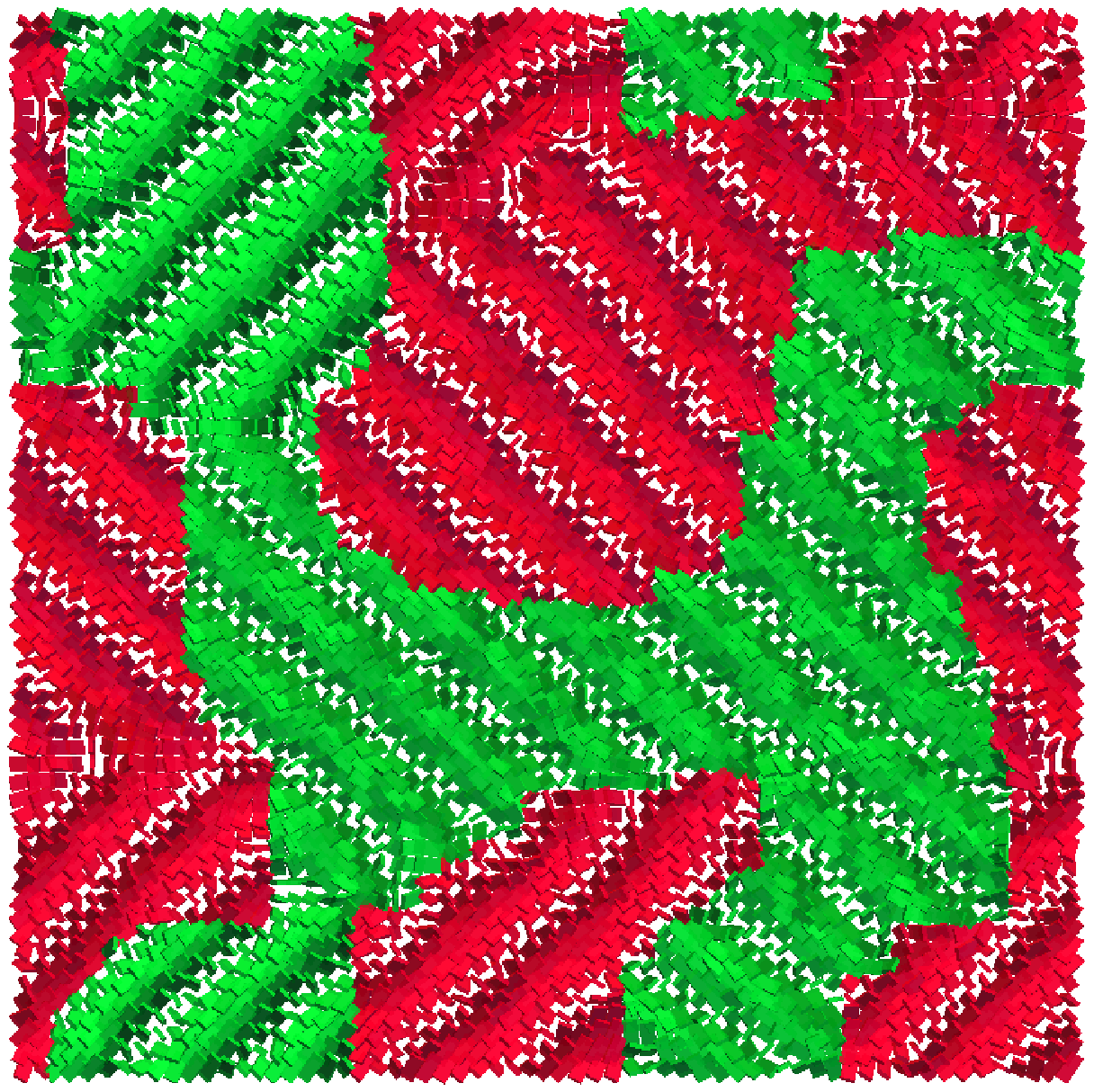}}
}
\caption{Exemplary snapshots of equilibrium configurations on $64 \times 64
$ lattice for $\lambda=0.3$, $\tau=\kappa=1$, and for different 
temperatures $t$, showing disordered and ordered structures. Sides of 
cuboids are moduli of eigenvalues of the $\mathbf{Q}$ tensor, while the 
color represents parity: $p=+1$ (red) and $p=-1$ (green). The phase 
transition temperature is $t^*=0.888 \pm 0.001$. Simulations were 
initialized from a random parity distribution and orientationaly 
disordered 
configuration of molecular tripods.
}
\label{fig:snapshots}
\end{figure}
Typical final configurations of the production run for $\lambda=0.3$ are 
shown in Fig.~\ref{fig:snapshots}, where
the transition between $I$ and $N^*_A$ is 
observed at $t^*=0.888 \pm 0.001$. As expected, for $t>t^*$ the isotropic 
structure is stabilized where small, orientationally ordered domains 
representing both chiralities are present. For temperatures slightly 
below the phase 
transition the system becomes orientationally ordered and chiral, with 
the main director perpendicular to the modulation axis. Because 
of 
a random initial state, chiral ambidextrous domains can form below $t^*$ 
as shown in Fig.~\ref{fig:snapshots}(d). 

\begin{figure}[htb]
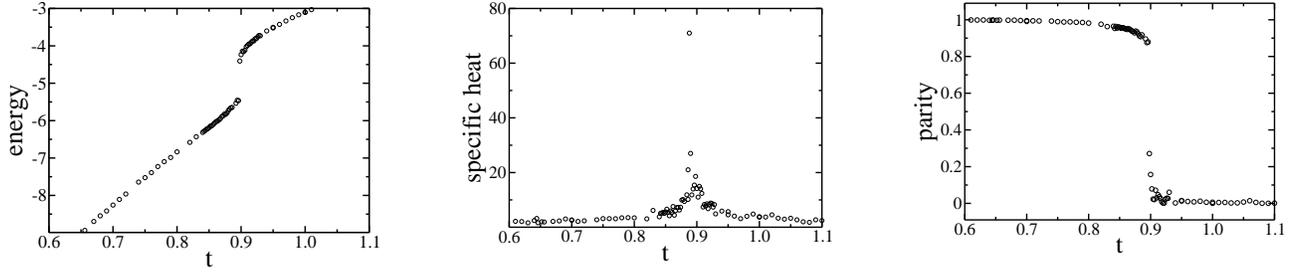

\centerline{%
\subfigure[]{\includegraphics[width=0.3\columnwidth]{e}}
\hspace{0.05\columnwidth}
\subfigure[]{\includegraphics[width=0.3\columnwidth]{cv}}
\hspace{0.05\columnwidth}
\subfigure[]{\includegraphics[width=0.3\columnwidth]{p}}
}
\caption{Temperature dependence of (a) energy, (b) specific heat, and (c) 
average parity for $\lambda=0.3$ and $\tau=\kappa=1$, calculated for 
equilibrium 
configurations on $32 \times 32$ lattice. Each dot correspond to an 
equilibrium 
value obtained from independent MC simulation started from random initial configurations. 
}
\label{fig:l03}
\end{figure}

\begin{figure}[htb]
\centerline{%
\includegraphics[width=0.5\columnwidth]{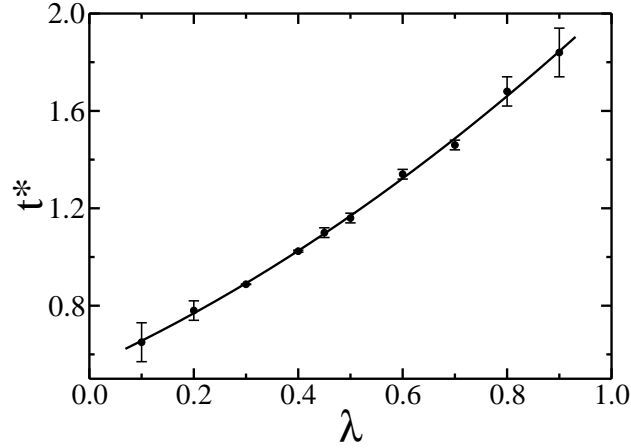}
}
\caption{Transition temperatures for $0.1\le\lambda\le 0.9$ 
estimated from a position of the peak in specific heat dependence on $t$. 
Error 
bars correspond to the width of the peak. The system's size is $32 \times 32$
and $\tau=\kappa=1$. Dots are data obtained from simulations; solid line is 
to guide the eye. 
}
\label{fig:tvl}
\end{figure}

The phase transition to $N^*_A$ is found by monitoring the 
average energy per molecule, the specific heat and the parity order parameter 
\begin{equation}\label{avparity}
\overline{p}= \overline{\frac{1}{N}  \sum_{i=1}^N p_i},
\end{equation}
where overline denotes thermodynamic average.
Exemplary temperature dependence of the average energy, the heat 
capacity and the parity order parameter
for $\lambda=0.3$ are shown in Fig.\ref{fig:l03}.
Similar plots are obtained both for $32 \times 32$ and $64 \times 64$ systems 
indicating that finite size effects are kept within the statistical error.
The average energy and the parity order 
parameter show a jump at $t^*\approx 0.888$
suggesting that the phase transition is of the first order. As the heat 
capacity is peaked around the same temperature the peak's position is 
used to determine the transition temperature.  With this 
identification of the phase transition temperature a  detailed  analysis of
$t^*$ as function of $\lambda$ is given in 
Fig.~\ref{fig:tvl}. It shows that $t^*$ increases 
with increasing  $\lambda$.

\begin{figure}[htb]
\centerline{%
\includegraphics[width=0.5\columnwidth]{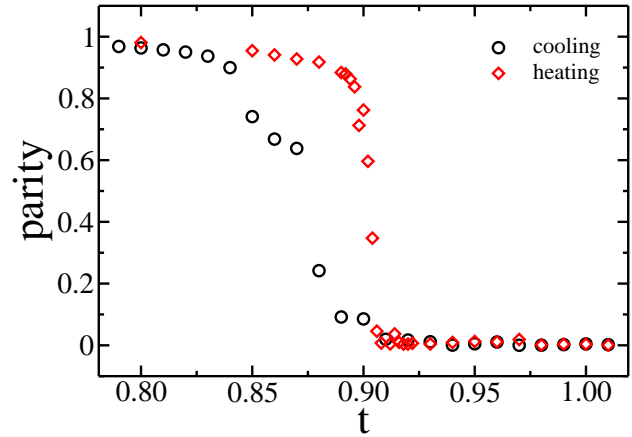}
}
\caption{Parity dependence on $t$ for $\lambda=0.3$, $\tau=\kappa=1$ when $64 
\times 64$ sample was heated (squares) and cooled (diamonds). Presence of 
considerable jump and hysteresis suggests that the transition is of the 
first 
order. 
}
\label{fig:chiralityl03}
\end{figure}

A further support for the first-order nature of the $I-N^*_A$ phase transition is 
the observation of hysteresis for the average parity (\ref{avparity}), as shown in 
Fig.~\ref{fig:chiralityl03}. More specifically, 
when the temperature scan, starting from a random initial configuration
in high temperature phase,
progresses by cooling down in small temperature steps to the final 
ordered state at low temperature and 
then is heated up until temperature of the high temperature state is reached 
again, the average parities obtained from the corresponding production runs on 
cooling and heating are slightly shifted. 

\begin{figure}[htb]
\centerline{%
\subfigure[]{\includegraphics[width=0.3\columnwidth]{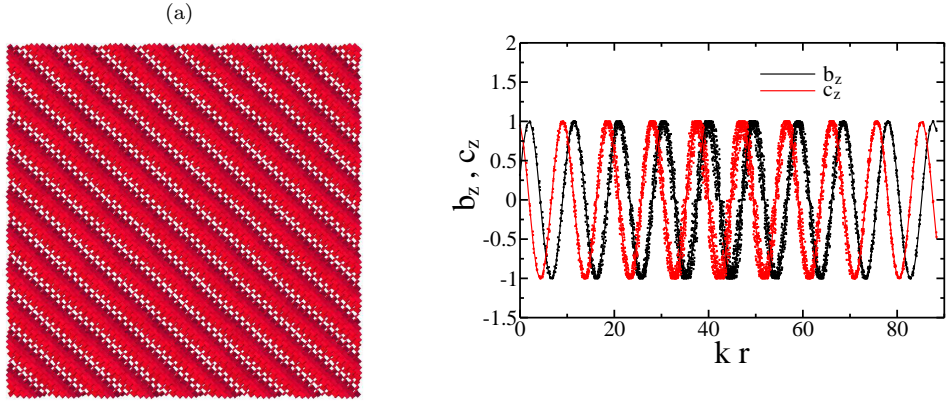}}
\hspace{0.05\columnwidth}
\subfigure[]{\includegraphics[width=0.4\columnwidth]{w2w3}}
}
\caption{Exemplary (a) snapshot of equilibrium  monodomain configuration
of parity $\overline{p}\approx +1$ for $\lambda=0.3$, $\tau=\kappa=1$ and
for very low temperature  $t=0.05$. 
Sides of 
cuboids are moduli of eigenvalues of the $\mathbf{Q}$ tensor. (b) The 
corresponding 
$z$-components of $\mathbf{\hat{b}}= \overline{\mathbf{\hat{b}}_i}$ and 
$\mathbf{\hat{c}}=\overline{\mathbf{\hat{c}}_i}$ 
 as function of 
position  $\mathbf{r}$ along the modulation axis $\mathbf
{k} \parallel \mathbf{\hat{a}} = [0.681, 0.732, 0.]$. Dots are data 
calculated 
for the snapshot (a) and lines are least-square fits: $b_{z}(\mathbf{k}
\cdot\mathbf{r}) = \sin(0.660 \, \mathbf{k}\cdot\mathbf{r} + 0.318)$ and 
$c_{z}(\mathbf{k}\cdot\mathbf{r}) = \sin(0.660 \, \mathbf{k}
\cdot\mathbf
{r} + 1.879)$.
}
\label{fig:w2w3}
\end{figure}

\begin{figure}[htb]
\centerline{%
\includegraphics[width=0.5\columnwidth]{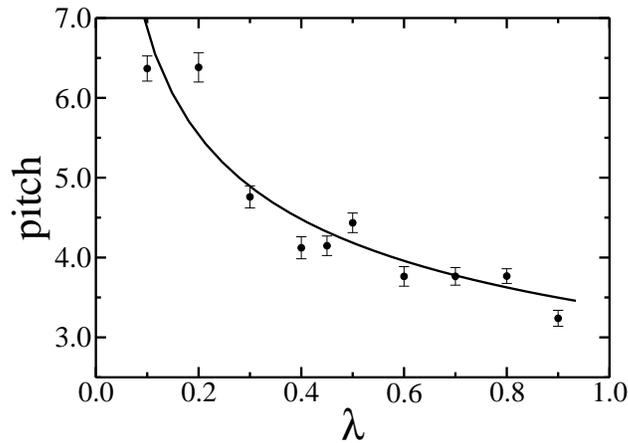}
}
\caption{Pitch of low temperature ambidextrous cholesteric phase  
as function of $\lambda$ for $\tau=\kappa=1$. Dots are data from simulations. The 
error bars are due to finite size of the system. Solid line is to guide the 
eye.
}
\label{fig:lambdadependence}
\end{figure}

To perform detailed structure analysis of low temperature phase we 
equilibriate monodomains of definite chirality, like the one shown in 
Fig.\ref{fig:w2w3}(a), and study average orientational properties of the molecular 
tripods $\{\mathbf{\hat{a}}_i, \mathbf{\hat{b}}_i, \mathbf{\hat{c}}_i\}$, Fig.\ref{fig:w2w3}(b).
Calculation of the average direction of these vectors shows that
only 
one of them,  $\mathbf{\hat{a}} = \overline{\mathbf{\hat{a}}_i}$, does not vanish.
The other two rotate around the $\mathbf{k}$-axis, which is parallel to 
$\mathbf{\hat{a}}$, Fig.\ref{fig:w2w3}(b). This behaviour is 
typical for cholesteric ordering with the cholesteric pitch, 
being in the studied case equal to $\pi/0.66 = 
4.76$.
Similar analysis can be carried out for cholesteric phases obtained for 
different values of $\lambda$. The corresponding cholesteric pitch  
is shown in Fig.\ref{fig:lambdadependence}. Note that the pitch decreases with 
$\lambda$.

Summarizing,  we showed that 
the coupling between quadrupolar and octupolar  interactions can
lead to chiral symmetry breaking in 2D with orientational 
arrangement similar to that observed for $N_{TB}$ in 3D. 
Although in two dimensions we do not expect any 
spontaneous breakdown of continuous symmetries 
the introduced molecular parity is a discrete, Ising-like molecular 
degree of freedom and the MCM hamiltonian (\ref{hamiltonian}) 
is $Z(2) \times SO(3)$ symmetric. Hence, it is not unreasonable to 
expect that a phase transition involving spontaneous parity (chirality) 
breaking can occur even in 2D. Indeed, as we demonstrated the new 
ambidextrous cholesteric phase can be stabilized from the 
isotropic phase through the first order phase transition
for the MCM 
model. This structure 
apparently relaxes the frustration, Fig.~\ref{fig:frustration}, of the 
hamiltonian`s ground state and is remarkably stable as seen from Fig~4(c).  

The results of simulations  indicate that the complex effects due
to spontaneous breaking of chirality, encountered \textit{e.g.} in bent-core 
and 
flexible dimer systems can be accounted for in a microscopic dispersion 
model 
with couplings such as (\ref{hamiltonian}).  Subsequent, detailed analysis 
needs to be undertaken to study the phase
transitions and remaining unidentified structures, which can exist in the 
three-dimensional case. Finally, because geometrical 
frustration and
chirality are known to lead to the emergence of blue phases, it is important 
to establish
whether a link between the model  (\ref{hamiltonian}) to those structures 
exists.

\section*{Acknowledgements}
This work was supported by Grant No. DEC-2013/11/B/ST3/04247 of the National Science Centre in Poland.

\bibliographystyle{tLCT}
\bibliography{main}

\end{document}